\documentclass[12pt,draft]{iopart}
\usepackage{amsfonts}
\usepackage[english]{babel}
\usepackage{euscript}
\newcommand{\textfrac}[2]{{\textstyle\frac{#1}{#2}}}

\begin{document}

\title[Symmetries and coverings for the r-th mdKP equation]{%
Cartan's structure of symmetry pseudo-group and 
\\
coverings for the r-th modified dispersionless 
\\
Kadomtsev--Petviashvili equation
}

\author{Oleg I. Morozov}

\address{Department of Mathematics, Moscow State Technical University 
\\
of Civil Aviation, Kronshtadtskiy Blvd 20, Moscow 125993, Russia
\\
oim{\symbol{64}}foxcub.org}


\begin{abstract}
We derive two non-equivalent coverings for the r-th mdKP equation from Maurer--Cartan forms of its symmetry pseudo-group. Also we find B\"acklund trans\-for\-ma\-ti\-ons between the covering equations.
\end{abstract}



\maketitle

\section{Introduction}

The role of coverings in studying nonlinear differential equations ({\sc de}s) is well-known, \cite{KV84,KLV,KV89,KV99}. They lead to a number of useful techniques such as inverse scattering trans\-for\-ma\-ti\-ons, B\"ack\-lund transformations, recursion operators, nonlocal symmetries and nonlocal conservation laws. For a given {\sc de}, a problem of constructing a covering is very difficult, see, e.g., 
\cite{WE,Estabrook,DoddFordy,Hoenselaers,Marvan1997,Marvan2002,Sakovich,Igonin,%
Morris1976,Morris1979,Zakharov82,Tondo,Marvan1992,Harrison1995,Harrison2000}. 
One of the pos\-sible approaches to solution lies in the framework of \'Elie Cartan's structure theory of Lie pseudo-groups, \cite{Kuzmina,BryantGriffiths,Morozov2007,Morozov2008}.

In the present paper we apply  the method of \cite{Morozov2007,Morozov2008} to the r-th modified dis\-per\-si\-on\-less Kadomtsev--Petviashvili equation (r-mdKP), \cite{Blaszak}. We use 
\'Elie Cartan's method of equivalence, \cite{Cartan,Gardner,Kamran,Olver95}, to compute Maurer--Cartan  ({\sc mc}) forms of the pseudo-group of contact symmetries of r-mdKP, and then find two linear combinations of these forms, whose horizontalizations provide covering equations of r-mdKP. Previosly this approach was applied in \cite{Morozov2008} to a particular case of r-mdKP -- modified dispersionless Kadomtsev--Petviashvili equation (mdKP), or modified  Khokhlov--Zabolotskaya equation, \cite{Kuzmina,Krichever,Kupershmidt}. Coverings for particular cases  of r-mdKP were found in
\cite{FerapontovKhusnutdinova,Dunajski,KonopelchenkoAlonso,ChangTu} via other methods.

\section{Preliminaries}
\subsection{Coverings of DEs}

Let $\pi_{\infty} : J^{\infty}(\pi) \rightarrow \mathbb{R}^n$ be the infinite jet bundle of local sections of the bundle $\pi : \mathbb{R}^n \times \mathbb{R} \rightarrow \mathbb{R}$. The coordinates on $J^{\infty}(\pi)$ are $(x^i, u_I)$, where
$I = (i_1,...,i_k)$ are symmetric multi-indices,  $i_1,...,i_k \in \{1,...,n\}$, $u_\emptyset= u$, and for any local section $f$ of $\pi$ there exists a section $j_{\infty}(f) : \mathbb{R}^n \rightarrow J^{\infty}(\pi)$ such that $u_I(j_{\infty}(f)) = \partial^{\# I}(f)/\partial x^{i_1} ... \partial x^{i_k}$, $\# I =\#(i_1,...,i_k) = k$.  The {\it total derivatives} on $J^{\infty}(\pi)$ are defined in the local coordintes as
\[
D_i = \frac{\partial}{\partial x^i}
+\sum \limits_{\# I \ge 0} u_{Ii} \, \frac{\partial}{\partial u_I}.
\]
\noindent
We have $[D_i, D_j] = 0$ for $i, j \in \{1,...,n\}$. A {\sc de} $F(x^i, u_K)=0$  defines a submanifold 
$\EuScript{E}^{\infty} = \{ D_I(F) =0 \,\,\vert\,\, \#I\ge 0\} \subset J^{\infty}(\pi)$,
where $D_I = D_{i_1}\circ ... \circ D_{i_k}$ for $I=(i_1,...,i_k)$. We denote restrictions of $D_i$ on $\EuScript{E}^{\infty}$ as $\overline{D}_i$.

In local coordinates, a {\it covering} over $\EuScript{E}^{\infty}$ is a bundle
$\widetilde{\EuScript{E}}^{\infty} = \EuScript{E}^{\infty} \times \EuScript{Q} \rightarrow \EuScript{E}^{\infty}$ with fibre coordinates $q^\alpha$, $\alpha \in \{1,..., N\}$ or $\alpha \in \mathbb{N}$, equipped with {\it ex\-ten\-ded total derivatives}
\[
\widetilde{D}_i = \overline{D}_i
+\sum \limits_{\alpha}
T^{\alpha}_i (x^j, u_I, q^{\beta})\,\frac{\partial}{\partial q^\alpha}
\]
such that $[\widetilde{D}_i, \widetilde{D}_j ]=0$ whenever $(x^i, u_I) \in \EuScript{E}^{\infty}$.

Dually, the covering is defined by the following differential 1-forms, \cite{WE},
\[
\omega^{\alpha} = d q^{\alpha}- T^{\alpha}_i (x^j, u_I, q^{\beta})\,dx^i
\]
\noindent
such that
$d \omega^{\alpha} \equiv 0 \,\,\,({\mathrm{mod}}\,\,\,\omega^{\beta}, \bar{\vartheta}_I )$ iff $(x^i,u_I) \in \EuScript{E}^{\infty}$, where $\bar{\vartheta}_I$ are restrictions of the contact forms 
$\vartheta_I = du_I-u_{I,k}\,dx^k$ on $\EuScript{E}^{\infty}$.

\subsection{Cartan's structure theory of contact symmetry pseudo-groups of DEs}

A {\it pseudo-group} on a manifold $M$ is a collection of local dif\-feo\-mor\-phisms of $M$, which is closed under composition {\it when defined}, contains an identity and is closed under inverse. A {\it Lie pseudo-group} is a pseudo-group whose diffeomorphisms are local analytic solutions of an involutive system of partial differential equations. \'Elie Cartan's approach to Lie pseudo-groups is based on a possibility to characterize transformations from a pseudo-group in terms of a set of invariant differential 1-forms called {\it Maurer--Cartan forms}. The {\sc mc} forms for a Lie pseudo-group can be computed by means of algebraic operations and differentiation. Expressions of differentials of the {\sc mc} forms in terms of themselves give {\it structure equations} of the pseudo-group. The structure equations contain the full information about their pseudo-group.

\vskip 5 pt
\noindent
EXAMPLE 1. 
Consider the bundle $J^2(\pi)$ of jets of the second order of the bundle $\pi$. A differential 1-form $\vartheta$ on $J^2(\pi)$ is called a {\it contact form} if it is annihilated by all 2-jets of local sections: $j_2(f)^{*}\vartheta = 0$. In the local coordinates every contact 1-form is a linear combination of the forms  $\vartheta_0 = du - u_{i}\,dx^i$, $\vartheta_i = du_i - u_{ij}\,dx^j$, $i, j \in \{1,...,n\}$, $u_{ji} = u_{ij}$. A local dif\-feo\-mor\-phism 
$\Delta : J^2(\pi) \rightarrow J^2(\pi)$, 
$\Delta : (x^i,u,u_i,u_{ij}) \mapsto (\overline{x}^i,\overline{u},\overline{u}_i,\overline{u}_{ij})$,
is called a {\it contact trans\-for\-ma\-tion} if for every contact 1-form $\overline{\vartheta}$ the form $\Delta^{*}\overline{\vartheta}$ is also contact. We denote by ${\rm{Cont}}(J^2(\pi))$  the pseudo-group  of contact transformations on $J^2(\pi)$.
Consider the following 1-forms 
\[
\Theta_0 = a\, \vartheta_0,
\quad
\Theta_i = g_i\,\Theta_0 + a\,B_i^k\,\vartheta_k,
\quad
\Xi^i =c^i\,\Theta_0+f^{ik}\,\Theta_k+b_k^i\,dx^k,
\]
\begin{equation}
\Theta_{ij} = a\,B^i_k\, B^j_l\,(du_{kl}-u_{klm}\,dx^m) + s_{ij}\,\Theta_0+w_{ij}^{k}\,\Theta_k+z_{ijk}\,\Xi^k ,
\label{LCF}
\end{equation}
\noindent
defined on $J^2(\pi)\times\EuScript{H}$, where $\EuScript{H}$ is an open subset of $\mathbb{R}^{(2 n+1)(n+3)(n+1)/3}$ with local coordinates $(a$, $b^i_k$, $c^i$, $f^{ik}$, $g_i$, $s_{ij}$, $w^k_{ij}$, $u_{ijk})$, 
$i,j,k \in \{1,...,n\}$, $i \le j$, such that $a\not =0$, $\det (b^i_k) \not = 0$, $f^{ik}=f^{ki}$,  $u_{ijk}=u_{ikj}=u_{jik}$, while $(B^i_k)$ is the inverse matrix for the matrix $(b^k_l)$.
As it is shown in \cite{Morozov2006}, the forms (\ref{LCF}) are {\sc mc} forms for ${\rm{Cont}}(J^2(\pi))$, that is, a local diffeomorphism
$\widehat{\Delta} : J^2(\pi) \times \EuScript{H} \rightarrow J^2(\pi) \times \EuScript{H}$
satisfies the conditions
$\widehat{\Delta}^{*}\, \overline{\Theta}_0 = \Theta_0$,
$\widehat{\Delta}^{*}\, \overline{\Theta}_i = \Theta_i$,
$\widehat{\Delta}^{*}\, \overline{\Xi}^i = \Xi^i$,
and $\widehat{\Delta}^{*}\, \overline{\Theta}_{ij} = \Theta_{ij}$
if and only if it is projectable on $J^2(\pi)$, and its projection
$\Delta : J^2(\pi) \rightarrow J^2(\pi)$ is a contact transformation.
The structure equations for ${\rm{Cont}}(J^2(\pi))$ have the form
\begin{eqnarray}
\fl
d \Theta_0 &=& \Phi^0_0 \wedge \Theta_0 + \Xi^i \wedge \Theta_i,
\nonumber
\\
\fl
d \Theta_i &=& \Phi^0_i \wedge \Theta_0 + \Phi^k_i \wedge \Theta_k
+ \Xi^k \wedge \Theta_{ik},
\nonumber
\\
\fl
d \Xi^i &=& \Phi^0_0 \wedge \Xi^i -\Phi^i_k \wedge \Xi^k
+\Psi^{i0} \wedge \Theta_0
+\Psi^{ik} \wedge \Theta_k,
\nonumber
\\
\fl
d \Theta_{ij} &=& \Phi^k_i \wedge \Theta_{kj} 
+ \Phi^k_j \wedge \Theta_{ki}
- \Phi^0_0 \wedge \Theta_{ij}
+ \Upsilon^0_{ij} \wedge \Theta_0
+ \Upsilon^k_{ij} \wedge \Theta_k + \Lambda_{ijk} \wedge \Xi^k,
\nonumber
\end{eqnarray}
where the additional forms $\Phi^0_0$, $\Phi^0_i$, $\Phi^k_i$, $\Psi^{i0}$, $\Psi^{ij}$,
$\Upsilon^0_{ij}$, $\Upsilon^k_{ij}$, and $\Lambda_{ijk}$ depend on differentials of the coordinates of $\EuScript{H}$.

\vskip 5 pt 
\noindent
EXAMPLE 2.
Suppose $\EuScript{E}$ is a second-order differential equation in one dependent and $n$ independent variables. We consider $\EuScript{E}$ as a submanifold in $J^2(\pi)$.
Let ${\rm{Cont}}(\EuScript{E})$ be the group of contact symmetries for $\EuScript{E}$. It consists of all the contact transformations on $J^2(\pi)$ mapping $\EuScript{E}$ to itself. Let 
$\iota_0 : \EuScript{E} \rightarrow J^2(\pi)$ be an embedding, and 
$\iota = \iota_0 \times \rm{id} : \EuScript{E}\times \EuScript{H} \rightarrow J^2(\pi)\times \EuScript{H}$. The {\sc mc} forms of ${\rm{Cont}}(\EuScript{E})$ can be derived from  the forms   $\theta_0 = \iota^{*} \Theta_0$,  $\theta_i= \iota^{*}\Theta_i$, 
$\xi^i = \iota^{*}\Xi^i$, and $\theta_{ij}=\iota^{*}\Theta_{ij}$ by means of Cartan's method of equivalence, see details and examples in \cite{FelsOlver,Morozov2002,Morozov2006}.

\section{Cartan's structure of the contact symmetry pseudo-group for r-mdKP}
The r-th mdKP 
\[
\fl
u_{tx} = -\frac{(3-r)\,(1-r)}{2}\,u_x^2 u_{xx} +\frac{r\,(3-r)}{2-r}\,u_x u_{xy} 
+\frac{3-r}{(2-r)^2}\,u_{yy}+\frac{(3-r)\,(1-r)}{2-r}\,u_y u_{xx},
\]
$r \in \mathbb{Z}\backslash \{2\}$,  was derived in \cite{Blaszak}. For a convenience of computations we use the following change of variables: 
\[
\tilde{t} = (3-r)\,t, 
\quad 
\tilde{x} = x,
\quad  
\tilde{y} = (2-r)\,y, 
\quad
\tilde{u} = -(1-r)\,u,
\]
where $r\not \in\{1, 2, 3\}$.  Then we have
\[
\tilde{u}_{\tilde{y}\tilde{y}} = \tilde{u}_{\tilde{t}\tilde{x}} +
\left(\frac{1}{2\,(1-r)}\,\tilde{u}_{\tilde{x}}^2+\tilde{u}_{\tilde{y}}\right)\,
\tilde{u}_{\tilde{x}\tilde{x}}+\frac{r}{1-r}\,\tilde{u}_{\tilde{x}}\,\tilde{u}_{\tilde{x}\tilde{y}}.
\]
We drop tildes and denote $\kappa = \frac{r}{1-r}$; this yields
\begin{equation}
u_{yy} = u_{tx}+\left(\frac{\kappa+1}{2}\,u_x^2+u_y\right)\,u_{xx} +\kappa\,u_x\,u_{xy}.
\label{main}
\end{equation}
The exceptional cases $r=2$ and $r=3$ correspond to $\kappa = -2$ and $\kappa = -\frac{3}{2}$, respectively. We will not consider the case of $r=1$. The case of $\kappa = -1$ is exceptional, too, since $r \rightarrow \infty$ when $\kappa \rightarrow -1$. In the cases of $\kappa = 0$, 
$\kappa =1$, and $\kappa=-1$ Eq. (\ref{main}) gets the forms of the mdKP equation, \cite{Kuzmina,Krichever,Kupershmidt},  
\begin{equation}
u_{yy} = u_{tx}+\left(\frac{1}{2}\,u_x^2+u_y\right)\,u_{xx},
\label{KZ}
\end{equation}
the dBKP equation, \cite{Takasaki,KonopelchenkoAlonso},
\begin{equation}
u_{yy} = u_{tx}+\left(u_x^2+u_y\right)\,u_{xx} +u_x\,u_{xy}.
\label{dBKP}
\end{equation}
and the equation describing Lorentzian hyper-CR Einstein--Weil structures, \cite{Dunajski,FerapontovKhusnutdinova},
\begin{equation}
u_{yy} = u_{tx}+u_y\,u_{xx} -u_x\,u_{xy}.
\label{Dunajski}
\end{equation}

We use the method outlined in the previous section to com\-pute {\sc mc} forms and struc\-ture equations of the pseudo-group of contact symmetries for Eq. (\ref{main}). The results de\-pend on $\kappa$.

When $\kappa \not \in \{-2,  -1\}$, the structure equations have the form
\begin{eqnarray}
\fl
d\theta_0&=&
\left(\eta_1+\xi^2-\textfrac{\kappa-4}{8}\,\xi^3-\textfrac{\kappa}{\kappa+2}\,\theta_{22}\right) 
\wedge \theta_0
+\xi^1 \wedge \theta_1
+\xi^2 \wedge \theta_2
+\xi^3 \wedge \theta_3,
\nonumber
\\
\fl
d\theta_1&=&
\left(
   \textfrac{3}{2}\,\eta_1
  -\textfrac{\kappa-4}{8}\,\xi^3
 - \textfrac{3(\kappa+1)}{\kappa+2}\,\theta_{22}   
\right) \wedge \theta_1 
+\left((\kappa+1)\,\theta_2+(\kappa+2)\,\xi^2\right) \wedge \theta_3
+\xi^1 \wedge \theta_{11}
\nonumber
\\
\fl
&&+\left(\textfrac{2\kappa^2+15\kappa+4}{8}\,\theta_2 
-\textfrac{\kappa^2-10\kappa+8}{8} \xi^2 
+\kappa\,\theta_{23}\right) \wedge \theta_0
+\xi^2 \wedge \theta_{12}
+\xi^3 \wedge \theta_{13},
\nonumber
\\
\fl
d\theta_2&=&\left(\textfrac{1}{2}\eta_1 
-\textfrac{\kappa+1}{\kappa+2}\, \theta_{22} 
+\textfrac{3\,\kappa+4}{8}\,\xi^3 \right)\wedge \theta_2
+\xi^1 \wedge \theta_{12}
+\xi^2 \wedge \theta_{22}
+\xi^3 \wedge \theta_{23},
\nonumber
\\
\fl
d\theta_3&=&
\left(\eta_1 -\textfrac{2(\kappa+1)}{\kappa+2}\,\theta_{22}
+\textfrac{\kappa+4}{8}\,\xi^3\right) \wedge \theta_3
+\left(\textfrac{\kappa-4}{8}\,\theta_{22} 
-\textfrac{\kappa^2-16}{64}\,\xi^3 \right)\wedge \theta_0
+\textfrac{\kappa+2}{2}\,\xi^2 \wedge \theta_2
\nonumber
\\
\fl
&&+\xi^1 \wedge \theta_{13} 
+\xi^2 \wedge \theta_{23}
+\xi^3 \wedge \theta_{12},
\nonumber
\\
\fl
d\xi^1&=&
-\left(\textfrac{1}{2}\,\eta_1  - \textfrac{2\,\kappa+3}{\kappa+2}\,\theta_{22}\right) \wedge \xi^1,
\nonumber
\\
\fl
d\xi^2&=&
\left(\textfrac{1}{2}\,\eta_1 +\textfrac{1}{\kappa+2}\,\theta_{22} +\xi^3\right) \wedge \xi^2
+\left(\textfrac{\kappa-4}{8}\,\theta_0-\theta_3\right) \wedge \xi^1 
-\theta_2 \wedge \xi^3,
\nonumber
\\
\fl
d\xi^3&=&-(\kappa+2)\,(\theta_2+\xi^2)\wedge \xi^1+\theta_{22} \wedge \xi^3,
\nonumber
\\
\fl
d\theta_{11}&=&
2\,\eta_1 \wedge \theta_{11} 
-\kappa\,\eta_2 \wedge \theta_{0}
+\eta_3 \wedge  \xi^2 
+\eta_4 \wedge \xi^3 
+ \eta_5 \wedge \xi^1 
+\kappa\,\left(\theta_3
+\textfrac{\kappa+8}{4}\,\theta_{12}\right) \wedge \theta_0
\nonumber
\\
\fl
&&+\left((4\,\kappa+3)\,\theta_{23}
-\textfrac{\kappa^2-18\,\kappa+20}{4} \,\xi^2 \right)\wedge \theta_1
-\left((2\,\kappa+3)\,\theta_{13}
+\textfrac{5\,\kappa^2+31\,\kappa+24}{4}\,\theta_1\right)\wedge \theta_2
\nonumber
\\
\fl
&&-\kappa\,\theta_3 \wedge \theta_{12}
-\left(\textfrac{5\,\kappa+6}{\kappa+2}\,\theta_{22} 
+ \textfrac{\kappa-4}{8}\,\xi^3\right) \wedge \theta_{11}
+(2\,\kappa+4)\,\xi^2 \wedge \theta_{13},
\nonumber
\\
\fl
d\theta_{12}&=&
\eta_1 \wedge \theta_{12}
+\eta_2 \wedge \xi^3 
+ \eta_3 \wedge \xi^1
+\theta_2 \wedge \left(\theta_{23}-\xi^2\right)
+\left(\theta_3-\textfrac{\kappa-4}{8}\,\theta_0
+\textfrac{3\,\kappa+4}{\kappa+2}\,\theta_{12}
\right) \wedge \theta_{22}
\nonumber
\\
\fl
&&+\textfrac{3\,\kappa+4}{8}\,\xi^3 \wedge \theta_{12},
\nonumber
\\
\fl
d\theta_{13}&=&
\case{3}{2}\,\eta_1 \wedge \theta_{13}
+\eta_2\wedge\xi^2 
+\eta_3 \wedge \xi^3 
+\eta_4 \wedge \xi^1
+
\textfrac{\kappa^3+10\,\kappa^2-32\,\kappa-96}{64}\,\theta_0
\wedge\xi^2
\nonumber
\\
\fl
&&-\left(\textfrac{\kappa^3+\kappa^2-14\,\kappa - 24}{16}\,\theta_2
+\textfrac{\kappa^2-3\,\kappa-4}{8}\,\theta_{23}
\right) \wedge \theta_0
+\left(\textfrac{\kappa-4}{8}\,\theta_{22} 
-\textfrac{\kappa^2-16}{64}\,\xi^3\right) \wedge \theta_1
\nonumber
\\
\fl
&&-\left(\textfrac{7\,\kappa^2+32\,\kappa+24}{8}\,\theta_3 
+(\kappa+2)\,\theta_{12}\right)\,\wedge \theta_2
+\left((2\,\kappa+1)\,\theta_{23} 
+\textfrac{\kappa^2+22\,\kappa+24}{8}\,\xi^2\right) \wedge \theta_3
\nonumber
\\
\fl
&&+\textfrac{3\,(\kappa+2)}{2}\,\xi^2\wedge\theta_{12}
-\left(\textfrac{4\,\kappa+5}{\kappa+2}\,\theta_{22} 
-\textfrac{\kappa+4}{8}\,\xi^3\right) \wedge \theta_{13},
\nonumber
\\
\fl
d\theta_{22}&=&
\left(\textfrac{3\,\kappa^2+18\,\kappa+24}{8}\,\theta_2 
+(\kappa+2)\,(\theta_{23}+\xi^2) \right)\wedge \xi^1
\nonumber
\\
\fl
d\theta_{23}&=&
\textfrac{1}{2}\,\eta_1 \wedge \theta_{23}
+\eta_2 \wedge \xi^1
-\left(\textfrac{3(\kappa+4)}{8}\,\theta_{22} 
+\textfrac{9\,\kappa^2+48\,\kappa+112}{64}\,\xi^3\right) \wedge \theta_2
+\left(\textfrac{3\,(\kappa+4)}{8}\,\theta_{23}-\xi^2\right)\wedge \xi^3
\nonumber
\\
\fl
&&
+\left(\textfrac{2\,\kappa+3}{\kappa+2}\,\theta_{23}-\textfrac{3\,\kappa-4}{8}\,\xi^2\right)
\wedge \theta_{22},
\nonumber
\\
\fl
d\eta_1&=&0,
\nonumber
\\
\fl
d\eta_2&=&
\eta_6 \wedge \xi^1 
+\left(\eta_1 +\textfrac{2\,(2\,\kappa+3)}{\kappa+2}\,\theta_{22} 
-\textfrac{3\,(\kappa+4)}{8}\,\xi^3\right)\wedge \eta_2
+\theta_3 \wedge \left(\textfrac{3\,(\kappa-4)}{8}\,\theta_{22}-\xi^3 \right)
\nonumber
\\
\fl
&&+\left(\textfrac{3\,\kappa^2-16\,\kappa+16}{64}\,\theta_{22} 
- \textfrac{\kappa-4}{8}\,\xi^3\right)\wedge \theta_0
+\left(\textfrac{3\,(\kappa+4)}{8}\,\theta_2 +\xi^2\right) \wedge \theta_{23}
\nonumber
\\
\fl
&&+\left(\textfrac{3\,(\kappa+4)}{8}\,\theta_{22}
+\textfrac{9\,\kappa^2+48\,\kappa+112}{64}\,  \xi^3\right)\wedge\theta_{12},
\nonumber
\\
\fl
d\eta_3&=&
\eta_6 \wedge \xi^3
+\eta_7 \wedge \xi^1
+\left((\kappa+3)\,\theta_2+(\kappa+2)\,\xi^2\right) \wedge \eta_2
+\left(\textfrac{\kappa-4}{8}\,\theta_1 -\theta_{13}\right) \wedge \theta_{22}
\nonumber
\\
\fl
&&+\left(\textfrac{3}{2}\,\eta_1 
+\textfrac{5\,\kappa+7}{\kappa+2}\,\theta_{22}
+\textfrac{3\,\kappa+4}{8}\,\xi^3\right)\wedge \eta_3
+(\kappa+2)\, \theta_3 \wedge (\theta_{23}+\xi^2)
\nonumber
\\
\fl
&&+\textfrac{\kappa-4}{64}\,\left((3\,\kappa^2+18\,\kappa+32)\,\theta_2
+8\,(\kappa+2)\,(\theta_{23}+\xi^2)\right)\wedge \theta_0
\nonumber
\\
\fl
&&+\left(\textfrac{3\,\kappa^2+18\,\kappa+32}{8}\,\theta_3
+\textfrac{6\,\kappa^2+29\,\kappa+28}{4}\,\theta_{12}\right)\wedge \theta_2
-\left(3\,(\kappa+1)\,\theta_{23}  -\textfrac{3\,\kappa^2+34\,\kappa+32}{8}\,\xi^2\right)\wedge \theta_{12},
\nonumber
\\
\fl
d\eta_4&=&
\eta_6 \wedge \xi^2
+\eta_7 \wedge \xi^3
+\eta_8 \wedge \xi^1
+\left(2\,\eta_1-\textfrac{2\,(3\,\kappa+4)}{\kappa+2}\,\theta_{22} 
+\textfrac{\kappa+4}{8}\,\xi^3 \right)\wedge \eta_4
\nonumber
\\
\fl
&&
+\kappa\,\left(\textfrac{\kappa-4}{8}\,\theta_0-2\,\theta_3\right)\wedge \eta_2
+\left((2\,\kappa+4)\,\theta_{2}+\textfrac{5\,(\kappa+2)}{2}\,\xi^2 \right)\wedge \eta_3
\nonumber
\\
\fl
&&-\textfrac{\kappa}{16}\,\left(2\,(\kappa-4)\,\theta_3
-(\kappa^2-2\,\kappa-4)\,\theta_{12}\right)\wedge \theta_0
+\textfrac{\kappa}{8}\,(7\,\kappa+5)\, \theta_3 \wedge \theta_{12}
\nonumber
\\
\fl
&&+\textfrac{1}{16}\,\left((2\,\kappa^3+3\,\kappa^2-30\,\kappa-112)\,\theta_2
-2\,(2\,\kappa^2-5\,\kappa+4)\,\theta_{23}\right)\wedge\theta_1
\nonumber
\\
\fl
&&-\textfrac{1}{32}\,(\kappa^3+10\,\kappa^2-32\,\kappa-96)\,\theta_1\wedge \xi^2
\nonumber
\\
\fl
&&
-\left(6\,(\kappa+1)\,\theta_{23}+\textfrac{1}{4}\,(\kappa^2+30\,\kappa+36)\,\xi^2\right)
\wedge \theta_{13}
\nonumber
\\
\fl
&&-\textfrac{1}{8}\,\left(20\,\kappa^2+101\,\kappa+92\right)\,\theta_2 \wedge \theta_{13}
-\textfrac{1}{64}\,\left(8\,(\kappa-4)\,\theta_{22}-(\kappa^2-16)\,\xi^3\right)\wedge\theta_{11},
\nonumber
\\
\fl
d\eta_5&=&
-\kappa\,\eta_6 \wedge \theta_0
+\eta_7 \wedge \xi^2
+\eta_8 \wedge \xi^3
+\eta_9 \wedge \xi^1
+\left(\textfrac{5}{2}\,\eta_1 
-\textfrac{7\,\kappa+9}{\kappa+2}\,\theta_{22}
-\textfrac{\kappa-4}{8}\,\xi^3 \right) \wedge \eta_5
\nonumber
\\
\fl
&&+(5\,\kappa+3)\,\eta_2 \wedge \theta_1
+\eta_3 \wedge \left(\textfrac{2\,\kappa^2+17\,\kappa-4}{8}\,\theta_0
+(\kappa-1)\,\theta_3\right)
-3\,(\kappa+1)\,\theta_{12} \wedge \theta_{13}.
\nonumber
\\
\fl
&&+\left((3\,\kappa+5)\,\theta_2 +3\,(\kappa+2)\,\xi^2 \right)\wedge \eta_4 
-\left(9\,(\kappa+1)\,\theta_{23}
-\textfrac{3\,\kappa^2-78\,\kappa-96}{8}\,\xi^2\right) \wedge \theta_{11}
\nonumber
\\
\fl
&&+\left(\textfrac{\kappa^3-22\,\kappa^2+52\,\kappa+90}{32}\,\theta_1
-\textfrac{\kappa^2+2\,\kappa-8}{4}\,\theta_{13}\right)\wedge \theta_0
-2\,(\kappa+2)\,\theta_3 \wedge \theta_{13}
\nonumber
\\
\fl
&&+\left(\textfrac{\kappa^2-22\,\kappa-20}{4}\,\theta_3
-\textfrac{6\,\kappa^2+39\,\kappa+24}{4}\,\theta_{12}\right) \wedge \theta_1
-\textfrac{24\,\kappa^2+71\,\kappa+64)}{4}\,\theta_2 \wedge \theta_{11}, 
\nonumber
\end{eqnarray}
where 
\begin{eqnarray}
\xi^1 &=& q^{-1}\,dt,
\nonumber
\\
\xi^2 &=& q\,u_{xx}^2\,\left(\left(\textfrac{\kappa+1}{2}\,u_x^2-u_y\right)\,dt + dx - u_x\,dy\right), 
\nonumber
\\
\xi^3 &=& u_{xx}\,\left(dy - (\kappa+2)\,u_x\,dt\right),
\label{MC_forms_general}
\\
\eta_1 &=& 2\,\frac{dq}{q}+\frac{2\,(2\,\kappa+3)}{\kappa+2}\,\frac{du_{xx}}{u_{xx}},
\nonumber
\end{eqnarray}
and $q=B_1^1 \not = 0$. We need not explicit expressions for the other {\sc mc} forms in the sequel.

\vskip 5 pt

In the case of $\kappa=-1$ the contact symmetry pseudo-group of Eq. (\ref{Dunajski}) has the 
fol\-low\-ing structure equations
\begin{eqnarray}
\fl
d\theta_0&=&\left(\eta_1 +\case{5}{8}\,\xi^3 +\theta_{22}\right) \wedge \theta_0
+\xi^1 \wedge \theta_1
+\xi^2 \wedge \theta_2
+\xi^3 \wedge \theta_3,
\nonumber
\\
\fl
d\theta_1&=&
\case{3}{2}\,\eta_1 \wedge \theta_1
+\eta_2 \wedge \left(\case{1}{8}\,\theta_0+\theta_3\right)
-\left(\case{9}{8}\,\theta_2 +\theta_{23} - \case{1}{2}\,\xi^2\right)\,\wedge\theta_0
-\case{5}{8}\,\theta_1 \wedge \xi^3 
+\xi^1 \wedge \theta_{11}
\nonumber
\\
\fl
&&+\xi^2 \wedge \theta_{12}
+\xi^3 \wedge \theta_{13},
\nonumber
\\
\fl
d\theta_2&=&\case{1}{8}\,\left(4\,\eta_1 +\xi^3\right) \wedge \theta_2
+\xi^1 \wedge \theta_{12}
+\xi^2 \wedge \theta_{22}
+\xi^3 \wedge \theta_{23},
\nonumber
\\
\fl
d\theta_3&=&
\eta_1 \wedge \theta_3
+\case{1}{2}\,\eta_2 \wedge \theta_2
-\case{5}{64}\,\left(8\,\theta_{22} +3\,\xi^3 \right)\wedge \theta_0
+\case{3}{8}\,\xi^3 \wedge \theta_3
+\xi^1 \wedge \theta_{13}
+\xi^2 \wedge \theta_{23}
\nonumber
\\
\fl
&&+\xi^3 \wedge \theta_{12},
\nonumber
\\
\fl
d\xi^1&=&-\case{1}{2}\,\left(\eta_1-2\,\theta_{22}\right) \wedge \xi^1,
\nonumber
\\
\fl
d\xi^2&=&
-\case{1}{8}\,\left(5\,\theta_0+8\,\theta_3\right) \wedge \xi^1
+\case{1}{2}\,\left(\eta_1 +2\,\theta_{22}+\xi^3\right) \wedge \xi^2
-\case{1}{2}\,\left(\eta_2 +\theta_2\right) \wedge \xi^3,
\nonumber
\\
\fl
d\xi^3&=&-\left(\eta_2+\theta_2\right) \wedge \xi^1+\theta_{22} \wedge \xi^3,
\nonumber
\\
\fl
d\theta_{11}&=&
2\,\eta_1 \wedge \theta_{11}
+\eta_2 \wedge \left(\case{3}{4}\,\theta_1+2\,\theta_{13}\right)
+\eta_4 \wedge \xi^2
+\eta_5 \wedge \xi^3
+\eta_6 \wedge \xi^1
-\left(\theta_{22} -\case{5}{8}\,\xi^3 \right) \wedge \theta_{11}
\nonumber
\\
\fl
&&+\left(\eta_3 -\theta_3-\case{7}{4}\,\theta_{1\,2}\right) \wedge \theta_0
-\left(\case{1}{2}\,\theta_2-\theta_{23}
-\case{1}{2}\,\xi^2 \right) \wedge \theta_1
+\theta_2 \wedge \theta_{13} 
+\theta_3 \wedge \theta_{12},
\nonumber
\\
\fl
d\theta_{12}&=&
\eta_1 \wedge \theta_{12}
+\eta_2 \wedge \left(\case{1}{8}\,\theta_2 +\theta_{23}+\xi^2\right)
+\eta_3 \wedge \xi^3 
+\eta_4 \wedge \xi^1
+\case{5}{8}\,\theta_0 \wedge \theta_{22}
-\xi^2 \wedge \theta_{23}
\nonumber
\\
\fl
&&-\left(\theta_{23}-\case{9}{8}\,\xi^2\right) \wedge \theta_2
+\left(\theta_3 +\theta_{12}\right)\wedge \theta_{22}
+\case{1}{8}\,\xi^3 \wedge \theta_{12},
\nonumber
\\
\fl
d\theta_{13}&=&
\case{3}{2}\,\eta_1 \wedge \theta_{13} 
+\case{1}{64}\,\eta_2 \wedge \left(35\,\theta_0+24\,\theta_3
+96\,\theta_{12}\right)
+\eta_3 \wedge \xi^2
+\eta_4 \wedge \xi^3
+\eta_5 \wedge \xi^1
\nonumber
\\
\fl
&&+\case{5}{16}\,\left(2\,\theta_2+\xi^2\right) \wedge \theta_0
-\case{5}{64}\,\left(8\,\theta_{22} -3\,\xi^3\right)\wedge \theta_1
+\left(\case{1}{8}\,\theta_3+\theta_{12}\right)\wedge \theta_2
+\theta_3 \wedge \theta_{23}
\nonumber
\\
\fl
&&-\left(\theta_{22} -\case{3}{8}\,\xi^3\right) \wedge \theta_{13},
\nonumber
\\
\fl
d\theta_{22}&=&
\case{1}{8}\,\left(4\,\eta_2 +9\,\theta_2 + 4\,\xi^2 +8\,\theta_{23}\right)\wedge \xi^1,
\nonumber
\\
\fl
d\theta_{23}&=&\case{1}{2}\,\eta_1 \wedge \theta_{23}
+\case{1}{2}\,\eta_2\wedge\left(\theta_{22}+\xi^3\right)
+\eta_3 \wedge \xi^1
+\case{1}{64}\,\theta_2\wedge \left(72\,\theta_{22}-73\,\xi^3\right)
\nonumber
\\
\fl
&&+\left(\theta_{23}+\case{3}{8}\,\xi^2 \right)\wedge \theta_{22}
+\case{1}{8}\,\left(9\,\theta_{23}+4\,\xi^2\right)\wedge \xi^3,
\nonumber
\\
\fl
d\eta_1&=&0,
\nonumber
\\
\fl
d\eta_2&=&
\case{1}{2}\,\left(\eta_1 +\xi^3\right) \wedge \eta_2
-\case{5}{8}\,\theta_0 \wedge \xi^1
-\left(\theta_2+\case{1}{2}\,\xi^2\right) \wedge \xi^3
-\theta_3 \wedge \xi^1
-\theta_{22} \wedge \xi^2,
\nonumber
\\
\fl
d\eta_3&=&
\eta_7 \wedge \xi^1
+\eta_1 \wedge \eta_3
+\case{1}{64}\,\eta_2 \wedge \left(41\,\theta_2 + 72\,\theta_{23}+36\,\xi^3\right)
+\case{1}{8}\,\eta_3 \wedge \left(16\,\theta_{22}+9\,\xi^3\right)
\nonumber
\\
\fl
&&+\case{5}{64}\,\left(7\,\theta_{22}+8\,\xi^3\right)\wedge \theta_0
-\case{1}{64}\,\left(72\,\theta_{23}+41\,\xi^2\right)\wedge \theta_2
-\left(\case{1}{8}\,\theta_{22}+\xi^3\right)\wedge \theta_3
\nonumber
\\
\fl
&&+\case{1}{64}\,\left(72\,\theta_{22}+73\,\xi^3\right)\wedge\theta_{12}
+\case{1}{8}\,\theta_{23}\wedge \xi^2,
\nonumber
\\
\fl
d\eta_4&=&
\eta_7 \wedge \xi^3
+\eta_8 \wedge \xi^1
+\case{1}{8}\,\left(5\,\theta_0+8\,\theta_3+6\,\theta_{12}\right)\wedge \eta_2
+\left(2\,\eta_2+2\,\theta_2-\xi^2\right)\wedge \eta_3
\nonumber
\\
\fl
&&+\case{1}{8}\,\left(12\,\eta_1-16\,\theta_{22}+\xi^3\right)\wedge \eta_4
+\case{5}{64}\,\theta_0\wedge\left(17\,\theta_2+8\,\theta_{23}\right)
-\case{5}{8}\,\theta_1 \wedge \theta_{22}
\nonumber
\\
\fl
&&+\case{1}{8}\,\left(17\,\theta_3+10\,\theta_{12}\right)\wedge\theta_2
+\theta_3 \wedge \theta_{23}
-\case{5}{8}\,\theta_{12} \wedge \xi^2 
-\theta_{13} \wedge \theta_{22},
\nonumber
\\
\fl
d\eta_5&=&
\eta_7 \wedge \xi^2
+\eta_8 \wedge \xi^3
+\eta_9 \wedge \xi^1
-\case{5}{32}\,\eta_2\wedge\left(7\,\theta_1+8\,\theta_{13}\right)
+\left(\case{5}{8}\,\theta_0+2\,\theta_3\right) \wedge \eta_3
\nonumber
\\
\fl
&&+\left(\case{5}{2}\,\eta_2 +2\,\theta_2\right) \wedge \eta_4 
+\left(2\,\eta_1 -2\,\theta_{22} +\case{3}{8}\,\xi^3\right)\wedge \eta_5
+\case{5}{16}\,\left(\theta_{12}-2\,\theta_3\right)\wedge\theta_0
\nonumber
\\
\fl
&&-\case{5}{16}\,\left(5\,\theta_2+2\,\theta_{23}+2\,\xi^2\right)\wedge\theta_1
+\case{13}{8}\,\theta_{12} \wedge \theta_3
-\case{5}{64}\,\left(8\,\theta_{22}+3\,\xi^3\right)\wedge\theta_{11}
\nonumber
\\
\fl
&&-\case{1}{8}\,\left(11\,\theta_2+4\,\xi^2\right)\wedge\theta_{13},
\nonumber
\\
\fl
d\eta_6&=&
\eta_7 \wedge \theta_0
+\eta_8 \wedge \xi^2
+\eta_9 \wedge \xi^3
+\eta_{10} \wedge \xi^1 
+\eta_2\wedge\left(\eta_5-\case{15}{8}\,\theta_{11}\right)
-2\,\eta_3 \wedge \theta_1
\nonumber
\\
\fl
&&-\eta_4\wedge\left(\case{19}{8}\,\theta_0+2\,\theta_3\right)
-2\,\eta_5 \wedge \theta_2
-\case{1}{8}\,\eta_6\wedge\left(20\,\eta_1+16\,\theta_{22}-5\,\xi^3\right)
\nonumber
\\
\fl
&&+\case{1}{32}\,\left(5\,\theta_1+72\,\theta_{13}\right)\wedge\theta_0
+\case{3}{4}\,\left(\theta_3+3\,\theta_{12}\right)\wedge\theta_1
+\case{5}{4}\,\theta_{11} \wedge \theta_2
+2\,\theta_{13} \wedge \theta_3
\nonumber
\end{eqnarray}
with
\begin{eqnarray}
\xi^1 &=& q^{-1}\,dt,
\nonumber
\\
\xi^2 &=& q\,u_{xx}^2 \left((s^2u_{xx}^2-s\,u_x u_{xx}-u_y)\,dt +dx - s\,u_{xx}\,dy\right),
\nonumber
\\
\xi^3 &=& u_{xx}\,\left((u_x - 2\,s\,u_{xx})\,dt +dy\right),
\label{MC_forms_exceptional_2}
\\
\eta_1 &=& 2 \,\frac{dq}{q} + 2\, \frac{du_{xx}}{u_{xx}},
\nonumber
\end{eqnarray}
where $s= B_3^2 \in \mathbb{R}$.

\section{Coverings of r-mdKP}

Following \cite{Morozov2007,Morozov2008}, we find linear combinations of the {\sc mc} forms 
(\ref{MC_forms_general}) and (\ref{MC_forms_exceptional_2}), which pro\-vide coverings of Eq. (\ref{main}) in the cases of $\kappa\not \in \{-2, -\frac{3}{2}, -1\}$, $\kappa = -\frac{3}{2}$, and $\kappa=-1$, res\-pec\-ti\-ve\-ly.
  
\subsection{General case}

When  $\kappa\not \in \{-2, -\frac{3}{2}, -1\}$, we take the following linear combination of 
the {\sc mc} forms (\ref{MC_forms_general})
\begin{eqnarray}
\omega &=& \eta_1 - \lambda_1\,\xi^1 - \lambda_2\,\xi^2 - \lambda_3\,\xi^3
\nonumber
\\
&&=2\,\frac{dq}{q}+\frac{2\,(2\,\kappa+3)}{\kappa+2}\,\frac{du_{xx}}{u_{xx}}
-\lambda_2\,q\,u_{xx}\,dx 
+\left(\lambda_2 q \,u_x u_{xx}-\lambda_3 u_{xx}\right)\,dy
\nonumber
\\
&&
+\left(\lambda_2q\,\left(u_y-\frac{\kappa+1}{2}\,u_x^2\right)\,u_{xx}^2
+\lambda_3\,(\kappa+2)\,u_xu_{xx}-\lambda_1 q^{-1}\right)\,dt, 
\nonumber
\end{eqnarray}
with $\lambda_1, \lambda_2, \lambda_3 \in \mathbb{R}$, and put
\[
q = -\frac{(\kappa+1)\,v}{(\kappa+2)\,v_1^{2\,\kappa+3}},
\qquad
u_{xx} = \frac{v_1^{\kappa+2}}{v},
\]
where $v$ and $v_1$ are new independent variables. This gives
\begin{equation}
\omega = -\frac{2\,(\kappa+1)}{(\kappa+2)\,v}\,
\left(dv - A\,v_1\,dt -v_1\,dx - B\,v_1\,dy\right)
\label{WE_form_general_1}
\end{equation}
with
\begin{eqnarray}
A &=& -\frac{(\kappa+2)^2}{4\,(\kappa+1)}\,\left(\lambda_1\lambda_2 v_1^{2(\kappa+1)}
+2\,\lambda_3\,(\kappa+1)\,u_x v_1^{\kappa+1}\right) 
+\frac{\kappa+1}{2}\,u_x^2-2\,u_y,
\nonumber
\\
B &=&-u_x -\frac{\lambda_3\,(\kappa+2)}{2\,(\kappa+1)}\,v_1^{\kappa+1}.
\nonumber
\end{eqnarray}
The form (\ref{WE_form_general_1}) is equal to zero whenever $v_1 = v_x$ and 
\begin{eqnarray}
v_t &=& \left(
\frac{\lambda_1\lambda_2\,(\kappa+2)^2}{4\,(\kappa+1)^2}\,v_x^{2(\kappa+1)}
+\frac{\lambda_3\,(\kappa+2)^2}{2\,(\kappa+1)}\,v_x^{\kappa+1}
+\frac{\kappa+1}{2} \,u_x^2-u_y
\right) v_x,
\nonumber
\\
v_y &=& -\left(\frac{\lambda_3\,(\kappa+2)}{\kappa+1}\,v_x^{\kappa+1}+u_x\right) v_x.
\nonumber
\end{eqnarray}
This system is compatible, i.e., $(v_t)_y = (v_y)_t$, whenever 
\begin{eqnarray}
u_{yy} - u_{tx} &-&\left(\frac{\kappa+1}{2}\,u_x^2+u_y\right)\,u_{xx} -\kappa\,u_x\,u_{xy}
\nonumber
\\
&&+
\frac{(\kappa+2)^2}{4\,(\kappa+1)^2}\,v_x^{2(\kappa+1)}\,
\left(\lambda_3^2\,(\kappa+2)^2- \lambda_1\lambda_2(2\,\kappa+3)\right) = 0.
\nonumber
\end{eqnarray}
This equation coincides with Eq. (\ref{main}) iff ~
$\lambda_3^2\,(\kappa+2)^2- \lambda_1\lambda_2(2\,\kappa+3) = 0$. So we put 
\[
\lambda_1 = \frac{\lambda_3^2\,(\kappa+2)^2}{\lambda_2\,(2\,\kappa+3)}.
\]
This yields
\begin{eqnarray}
\fl
v_t &=& \left(
\frac{\lambda_3^2\,(\kappa+2)^4}{(2\,\kappa+3)\,(\kappa+1)^2}\,v_x^{2(\kappa+1)}
+\frac{\lambda_3(\kappa+2)^2}{2\,(\kappa+1)}\,u_x\,v_x^{\kappa+1}
+\frac{\kappa+1}{2} \,u_x^2-u_y
\right) v_x,
\nonumber
\\
\fl
v_y &=& -\left(\frac{\lambda_3(\kappa+2)}{\kappa+1}\,v_x^{\kappa+1}+u_x\right) v_x.
\nonumber
\end{eqnarray}
When $\lambda_3 = 0$, we have 
\begin{equation}
v_t = \left(\frac{\kappa+1}{2} \,u_x^2-u_y\right) v_x,
\qquad
v_y = -u_x\, v_x. 
\label{covering_general_1}
\end{equation}
In the case of $\kappa = 0$ this covering for Eq. (\ref{KZ}) was obtained in \cite{Morozov2008}. 
When $\lambda_3 \not = 0$, we put 
$v = \left(\frac{\lambda_3\,(\kappa+2)}{2\,(\kappa+1)}\right)^{1/(\kappa+1)}\,w$. Then
\begin{eqnarray}
w_t &=& \left(
\frac{(\kappa+2)^2}{2\,\kappa+3}\,w_x^{2(\kappa+1)}
+(\kappa+2)\,u_x\,w_x^{\kappa+1}+\frac{\kappa+1}{2} \,u_x^2-u_y\right) w_x,
\nonumber
\\
w_y &=& -\left(w_x^{\kappa+1}+u_x\right) w_x.
\label{covering_general_2}
\end{eqnarray}
For $\kappa= 0$ this covering of (\ref{KZ}) was found in \cite{ChangTu} by means of another technique and in \cite{Morozov2008} via the method  described above. For $\kappa= 1$  the covering (\ref{covering_general_2}) of Eq. (\ref{dBKP}) was obtained in \cite{KonopelchenkoAlonso}.  

From  (\ref{covering_general_1}) we have
\begin{equation}
u_x = -\frac{v_y}{v_x},
\qquad
u_y = \frac{\kappa+1}{2}\,\left(\frac{v_y}{v_x}\right)^2 - \frac{v_t}{v_x}.
\label{covering_general_1_inverse}
\end{equation}
The integrability condition $(u_x)_y=(u_y)_x$ of this system gives  
\begin{equation}
v_{yy} = v_{tx} +\left(\frac{(\kappa+1)\,v_y^2}{v_x^2}-\frac{v_t}{v_x}\right)\,v_{xx}  
- \frac{\kappa\,v_y}{v_x}\,v_{xy}.
\label{covering_equation_general_1}
\end{equation}
For $\kappa = 0$ this equation was obtained in \cite{BogdanovKonopelchenko}.
Also, from (\ref{covering_general_2}) we get  
\begin{equation}
\fl
u_x = -\frac{w_y}{w_x}-w_x^{\kappa+1},
\quad
u_y  =-\frac{w_t}{w_x}+\frac{(\kappa+1)\,w_y^2}{2\,w_x^2}-w_x^\kappa w_y
-\frac{(\kappa+1)\,w_x^{2(\kappa+1)}}{2(2\kappa+3)}
\label{covering_general_2_inverse}
\end{equation}
This system yields
\begin{eqnarray}
\fl
w_{yy} 
&=&
w_{tx} 
+ \left((\kappa+1)\,\frac{w_y^2}{w_x^2}-\frac{w_t}{w_x}+\kappa\,w_x^\kappa w_y 
+\textfrac{(\kappa+1)^2}{2\,\kappa+3}\,w_x^{2(\kappa+1)}
\right)\, w_{xx}
\nonumber
\\
\fl 
&&
-\kappa\,\left(\frac{w_y}{w_x}+\,w_x^{\kappa}\right)\,w_{xy}.
\label{covering_equation_general_2}
\end{eqnarray}
Substitution for (\ref{covering_general_1_inverse}) in (\ref{covering_general_2}) gives a B\"acklund trnasformation 
\[
w_t = \frac{(\kappa+2)^2}{2\,\kappa+3}\,w_x^{2\kappa+3} 
-\frac{(\kappa+2) v_y}{v_x}\,w_x^{\kappa+2} 
+\frac{v_t}{v_x}\,w_x,
\quad
w_y =-w_x^{\kappa+2}+\frac{v_y}{v_x}\,w_x 
\]
from Eq. (\ref{covering_equation_general_1}) to 
Eq. (\ref{covering_equation_general_2}). The inverse B\"acklund transformation appears from sub\-sti\-tu\-ti\-on for (\ref{covering_general_2_inverse}) in (\ref{covering_general_1}).

\subsection{Case of $\kappa = -\frac{3}{2}$}

In the case of $\kappa=-\frac{3}{2}$  we take the following combination of the {\sc mc} forms (\ref{MC_forms_general}) 
\[
\fl
\omega = \eta_1 -\lambda_1\,\xi^1 - 4\, \xi^2 
= 2\,\frac{dq}{q} 
+\left(q\,u_{xx}^2\,(u_x^2+4\,u_y)-\lambda_1\,q^{-1}\right)\,dt - 4\,q\,u_{xx}\,\left(dx - u_x\,dy\right)
\]
and the following change of variables: $q=-v^{-2}$, $u_{xx} = (v\,v_1)^{1/2}$. Then we have
\[
\omega = - 4\,\frac{dv}{v} 
-\left(\frac{(u_x^2+4\,u_y)\,v_1}{v}-\lambda_1\,v^2\right)\,dt 
+\frac{4\,v_1}{v}\,dx - \frac{4\,u_x\,v_1}{v}\,dy.
\]
This form is equal to zero whenever $v_1= v_x$ and
\[
v_t = \case{1}{4}\,\lambda_1\,v^3 - \left(\case{1}{4}\,u_x^2+u_y\right)\,v_x,
\qquad
v_y = -u_x\,v_x.
\]
This system is compatible for every value of $\lambda_1$ whenever Eq. (\ref{main}) with 
$\kappa = -\frac{3}{2}$ is sa\-tis\-fi\-ed. When $\lambda_1=0$, we have Eqs. (\ref{covering_general_1}) with $\kappa = -\frac{3}{2}$:
\begin{equation}
v_t = -\left(\case{1}{4}\,u_x^2+u_y\right)\,v_x,
\qquad
v_y = -u_x\,v_x.
\label{first_covering_exceptional_1}
\end{equation}
When $\lambda_1 \not = 0$, we put $v=2\,\lambda_1^{-1/2}\,w$. Then we get
\begin{equation}
w_t = w^3-\left(\case{1}{4}\,u_x^2+u_y\right)\,w_x,
\qquad
w_y = -u_x\,w_x.
\label{second_covering_exceptional_1}
\end{equation}
Exclusion of  $u_x$ and $u_y$ from Eqs. (\ref{first_covering_exceptional_1}) and 
(\ref{second_covering_exceptional_1}) gives  equations
\begin{eqnarray}
v_{yy} &=& v_{tx} -\left(\frac{v_y^2}{2\,v_x^2}+\frac{v_t}{v_x}\right)\,v_{xx}  
+ \frac{3\,v_y}{2\,v_x}\,v_{xy},
\label{first_covering_equation_exceptional_1}
\\
w_{yy} &=& w_{tx} -\left(\frac{w_y^2}{2\,w_x^2}+\frac{w_t}{w_x}-\frac{w^3}{w_x}\right)\,w_{xx}  
+ \frac{3\,w_y}{2\,w_x}\,w_{xy} - 3\,w^2\,w_x,
\label{second_covering_equation_exceptional_1}
\end{eqnarray}
and a B\"acklund transformation from (\ref{first_covering_equation_exceptional_1})
to (\ref{second_covering_equation_exceptional_1}):
\[
w_t = w^3 +\frac{v_t}{v_x}\,w_x,
\qquad
w_y = \frac{v_y}{v_x}\,w_x.
\]

\subsection{Case of $\kappa = -1$}

When $\kappa= -1$, we take the following combination of the {\sc mc} forms (\ref{MC_forms_exceptional_2}): 
\begin{eqnarray}
\fl
\omega &=& \eta_1 +\case{1}{2}\,\lambda_3^2\,\xi^1+2\,\xi^2-\lambda_3\,\xi^3
=
2\,q\,u_{xx}^2\,dx 
- u_{xx}\left(2\,q\,s\,u_{xx}^2+\lambda_3\right)\,dy
+2\,\frac{dq}{q}+2\,\frac{du_{xx}}{u_{xx}} 
\nonumber
\\
\fl
&&
+ \frac{1}{2\,q}\,\left(\left(\lambda_3+q\,u_{xx}(2\,s\,u_{xx}-u_x\right)^2
-q^2u_{xx}^2\,\left(u_x^2+4\,u_y\right)\right)\,dt.
\nonumber
\end{eqnarray}
Then we substitute for $q=(v\,v_1)^{-1}$,
$s = u_x\,v_1^{-1}$, $u_{xx}= v_1$ and obtain
\[
\fl
\omega = -\frac{1}{2\,v}\,\left(4\,dv 
-\left(\lambda_3^2\,v^2+2\,\lambda_3\,u_x\,v-4\,u_y\right)\,v_1\,dt-4\,v_1\,dx 
+2\,\left(\lambda_3\,v+2\,u_x\right)\,v_1\,dy\right).
\]
This form is equal to zero whenever $v_1 = v_x$ and 
\[
v_t  =\left(\frac{\lambda_3^2}{4}\,v^2+\frac{1}{2}\,u_x\,v-u_y\right)\,v_x,
\qquad
v_y = -\left(\frac{1}{2}\,\lambda_3\,v+u_x\right)\,v_x.
\]
This system is compatible for every value of $\lambda_3$ whenever  Eq. (\ref{Dunajski}) is satisfied.  For $\lambda_3 = 0$ we have 
\begin{equation}
v_t  =-u_y\,v_x,
\qquad
v_y = -u_x\,v_x.
\label{first_covering_exceptional_2}
\end{equation}
When $\lambda_3 \not = 0$, we put $v = 2\,\lambda_3^{-1}\,w$. Then we have 
\begin{equation}
w_t = \left(w^2+u_x\,w-u_y\right)\,w_x,
\qquad
w_y = -(w+u_x)\,w_x.
\label{second_covering_exceptional_2}
\end{equation}
Excluding $u_x$ and $u_y$ from systems (\ref{first_covering_exceptional_2}) and 
(\ref{second_covering_exceptional_2}), we get equations 
\begin{equation}
v_{yy} 
= 
v_{tx} -\frac{v_t}{v_x}\,v_{xx}+\frac{v_y}{v_x}\,v_{xy}
\label{first_covering_equation_exceptional_2}
\end{equation}
and
\begin{equation}
w_{yy} 
=
w_{tx} -\frac{w_t+w\,w_x}{w_x}\,w_{xx}+\frac{w_y+w\,w_x}{w_x}\,w_{xy},
\label{second_covering_equation_exceptional_2}
\end{equation}
correspondingly. The B\"acklund transformation between Eqs. (\ref{first_covering_equation_exceptional_2}) and (\ref{second_covering_equation_exceptional_2})
has the form
\[
w_t = \left(w^2+\frac{v_t-w\,v_y}{v_x}\right)\,w_x,
\qquad
w_y = \left(w-\frac{v_y}{v_x}\right)\,w_x.
\]

\vskip 5 pt
\noindent
REMARK 1.  
A one-parametric family of coverings with a nonremovable parameter
\begin{equation}
v_t = -(u_y -\lambda\,u_x - \lambda^2)\,v_x,
\qquad
v_y = -(u_x+\lambda)\,v_x,
\label{Dunajski_covering}
\end{equation}  
for Eq. (\ref{Dunajski}) is presented in \cite{Dunajski}. This family can be obtained from 
(\ref{first_covering_exceptional_2})  by means the fol\-low\-ing technique, \cite[\S~3.6]{KV89}
\cite{Krasilshchik2000,IgoninKrasilshchik,Marvan2002,IgoninKerstenKrasilshchik}. 
Eq. (\ref{Dunajski}) has the infinitesimal symmetry 
\[
X = y\,\frac{\partial}{\partial x}+2\,x\,\frac{\partial}{\partial u}, 
\]
which can't be lifted into a symmerty of the covering (\ref{first_covering_exceptional_2}). Then the deformation 
$e^{\lambda X}$ transforms the covering (\ref{first_covering_exceptional_2}) into (\ref{Dunajski_covering}). Indeed, we have
\begin{eqnarray}
\tilde{t} &=& e^{\lambda\,X}(t) = t,
\quad 
\tilde{x} = e^{\lambda\,X}(x) = x+\lambda\,y,
\quad
\tilde{y} = e^{\lambda\,X}(y) = y,
\nonumber
\\
\tilde{u} &=& e^{\lambda\,X}(u) = u+2\,\lambda\,x+\lambda^2\,y,
\nonumber
\end{eqnarray}
and therefore
\[
\fl
\tilde{u}_{\tilde{t}} = e^{\lambda\,X}(u_t) = u_t,
\quad
\tilde{u}_{\tilde{x}} = e^{\lambda\,X}(u_x) = u_x+2\,\lambda,
\quad
\tilde{u}_{\tilde{y}} = e^{\lambda\,X}(u_y) = u_y-\lambda\,u_x-\lambda^2.
\] 
Since $\tilde{v} = e^{\lambda\,X}(v) = v$,  for the form 
\[
\tilde{\omega}_1 = d\tilde{v} + \tilde{u}_{\tilde{y}}\,\tilde{v}_{\tilde{x}}\,d\tilde{t} 
- \tilde{v}_{\tilde{x}}\,d\tilde{x} +\tilde{u}_{\tilde{x}}\,\tilde{v}_{\tilde{x}}\,d\tilde{y},
\]
which defines the covering (\ref{first_covering_exceptional_2}) in the tilded variables, 
we have
\[
\left(e^{\lambda X}\right)^{*} \tilde{\omega}_1 = 
dv + (u_y-\lambda\,u_x-\lambda^2)\,v_x\,dt - v_x\,dx +(u_x+\lambda)\,v_x\,dy.
\]
This form defines the family of coverings (\ref{Dunajski_covering}).

Similarly, we derive  a new one-parametric family of coverings with a non\-re\-mo\-vable parameter from the covering (\ref{second_covering_exceptional_2}). We have 
$\tilde{w} = e^{\lambda\,X}(w) = w$,  so the form 
\[
\tilde{\omega}_2 = d\tilde{w} 
-\left(\tilde{w}^2+\tilde{u}_{\tilde{x}}\,\tilde{w}
- \tilde{u}_{\tilde{y}}\right)\,\tilde{w}_{\tilde{x}}\,d\tilde{t} 
- \tilde{w}_{\tilde{x}}\,d\tilde{x} 
+\left(\tilde{w}+\tilde{u}_{\tilde{x}}\right)\,\tilde{w}_{\tilde{x}}\,d\tilde{y},
\]
which defines the covering (\ref{second_covering_exceptional_2}) in the tilded variables, provides
\begin{eqnarray}
\left(e^{\lambda X}\right)^{*} \tilde{\omega}_2 &=& 
dw -\left(w^2+(u_x+2\,\lambda)\,w- u_y+\lambda\,u_x+\lambda^2\right)\,w_x\,dt 
- w_x\,dx 
\nonumber
\\
&&+(w+u_x+\lambda)\,w_x\,dy.
\nonumber
\end{eqnarray}
This form defines a family of coverings 
\[
\fl
w_t = \left(w^2+(u_x+2\,\lambda)\,w- u_y+\lambda\,u_x+\lambda^2\right)\,w_x,
\qquad
w_y = -(w+u_x+\lambda)\,w_x.
\]

\vskip 5 pt
\noindent 
REMARK 2. In the case of $\kappa = -2$ Eqs. (\ref{covering_general_1}) define a covering for Eq. (\ref{main}), too, while we can't obtain this result by the method described above.

\end{document}